\documentclass{article}

\usepackage{arxiv}
\usepackage[utf8]{inputenc} 
\usepackage[T1]{fontenc}    
\usepackage{hyperref}       
\usepackage{url}            
\usepackage{booktabs}       
\usepackage{amsfonts}       
\usepackage{nicefrac}       
\usepackage{microtype}      
\usepackage{lipsum}
\usepackage{amssymb,amsmath,amsfonts,amstext,geometry,wasysym,marvosym,natbib,bm,graphicx,graphics,epsfig}
\usepackage{cleveref}

\title{Fitting IVIM with Variable Projection and Simplicial Optimization}

\author{
  Shreyas Fadnavis\thanks{corresponding author: shfadn@indiana.edu} \\
  Intelligent Systems Engineering\\
  Indiana University Bloomington\\
  \And
  Hamza Farooq \\
  Center for Magnetic Resonance Research\\
  University of Minnesota Twin Cities\\
  \And
  Maryam Afzali\\
  Cardiff University Brain Research Imaging Center\\
  Cardiff University\\
  \And
  Christoph Lenglet\\
  Center for Magnetic Resonance Research\\
  University of Minnesota Twin Cities\\
  \And
  Tryphon Georgiou\\
  Mechanical and Aerospace Engineering\\
  University of California, Irvine
  \And
  Hu Cheng\\
  Psychology and Brain Sciences\\
  Indiana University Bloomington\\
  \And
  Sharlene Newman\\
  Psychology and Brain Sciences\\
  Indiana University Bloomington\\
  University of Alabama\\
  \And
  Shahnawaz Ahmed\\
  Applied Quantum Physics Laboratory\\
  Chalmers University of Technology\\
  \And
  Rafael Neto Henriques\\
  Center for the Unknown\\
  Champalimaud Foundation\\
  \And
  Eric Peterson\\
  HeartVista, Inc.\\
  \And
  Serge Koudoro\\
  Intelligent Systems Engineering\\
  Indiana University Bloomington\\
  \And
  Ariel Rokem\\
  eScience Institute\\
  University of Washington\\
  \And
  Eleftherios Garyfallidis\\
  Intelligent Systems Engineering\\
  Indiana University Bloomington\\
}

\begin{document}
\maketitle

\begin{abstract}
Fitting multi-exponential models to Diffusion MRI (dMRI) data has always been challenging due to various underlying complexities. In this work, we introduce a novel and robust fitting framework for the standard two-compartment IVIM microstructural model. This framework provides a significant improvement over the existing methods and helps estimate the associated  diffusion and perfusion parameters of IVIM in an automatic manner. As a part of this work we provide capabilities to switch between more advanced global optimization methods such as simplicial homology (SH) and differential evolution (DE). Our  experiments show that the results obtained from this simultaneous fitting procedure disentangle the model parameters in a reduced subspace. The proposed framework extends the seminal work originated in the MIX framework, with improved procedures for multi-stage fitting. This framework has been made available as an open-source Python implementation and disseminated to the community through the DIPY project.
\end{abstract}


\section{Introduction}
The Intravoxel Incoherent Motion (IVIM) model uses data collected with the pulsed gradient spin echo sequence, often used to measure diffusion-weighted MRI contrasts, to derive information about blood micro-circulation. In IVIM, information about perfusion of blood is estimated as a pseudo-diffusion process from a PGSE experiment, in which several low b-values are added to sensitize the measurement both to diffusion, as  well as to perfusion \citep{bihan_2019}. The IVIM model has found important applications in a wide variety of perfusion-driven imaging methods since its introduction  \citep{bihan_breton_lallemand_aubin_vignaud_laval-jeantet_1988}. For example, IVIM is applicable to the physiology of brain tumors \citep{federau_meuli_obrien_maeder_hagmann_2013} along with monitoring the response of treatment to these tumors \citep{kim_suh_kim_choi_kim_2013}. Over the past twenty years, IVIM has shown promising results in imaging of breast \citep{iima_kataoka_2018}, nasopharyngeal \citep{lai_lee_lam_huang_chan_khong_2017}, pancreatic \citep{kang_lee_yoon_kiefer_han_choi_2013}, and liver \citep{li_cercueil_yuan_chen_loffroy_wang_2017} tumors. In addition, the IVIM model can be used in virtual MR elastography \citep{bihan_ichikawa_motosugi_2017} and MR angiography \citep{bihan_2019}.
 
 The IVIM signal model is expressed as a mixture of two exponential components, representing the standard bi-compartmental isotropic signal decay model for perfusion and diffusion. This makes it  difficult to fit, for the following reasons: (1) The exponential components in the signal model are non-orthogonal, which makes it hard to project it along the real-axis by taking an integral and (2) The inherent resolution limit in modeling the exponential decay of multi-exponential models  \citep{istratov_vyvenko_1999}. The present work tackles these challenges by simultaneously fitting the model parameters without an empirical fixed threshold on the b-values. In previous work, a variety of different approaches were taken to fitting the IVIM model \citep{jalnefjord_andersson_montelius_starck_elf_johanson_svensson_ljungberg_2018}. However, there has been no consensus about the best method to use \citep{bihan_2019}. Furthermore, the existing literature on fitting methods underscores the trade-off between precision and bias in  model estimation.
 
In the signal equation of this model, two exponentials are entangled: their linear coefficients represent volume fractions of a diffusion-weighted component and a perfusion-weighted component, and their exponents represent the apparent diffusion coefficient (ADC) and a pseudo-diffusion coefficient of these components \citep{bihan_breton_lallemand_aubin_vignaud_laval-jeantet_1988}. Fitting this model can be solved as a simple least squares problem, by minimizing the $L_2$ norm of the residuals between the observed and the predicted values. Unfortunately, because of the challenges mentioned above, fitting the linear coefficients and the exponents simultaneously is a hard nonlinear problem. Several methods have been proposed to analyze and fit multi-exponential functions, such as graphical analysis (peeling method) \citep{bell_1965} \citep{ferrante_ottenstein_warren_1987}, Nonlinear Least Squares (NLS) \citep{byrd_schnabel_shultz_1988}, algebraic techniques (Prony’s method) \citep{osborne_smyth_1995} and Pade-Laplace algorithm \citep{halvorson_1992}.

Here, we propose to use an approach previously used in the MIX framework \citep{farooq_xu_nam_keefe_yacoub_georgiou_lenglet_2016} by improving the global optimization strategy. The MIX framework showed that the Variable Projection approach was beneficial to fit multi-compartment microstructure models for crossings. For the IVIM model at hand, it is particularly useful to separate the treatment of linear parameters (i.e. the perfusion and diffusion fractions) from that of the nonlinear parameters (i.e. the perfusion and diffusion coefficients) via the Variable Projection method \citep{golub_pereyra_2003}. The main advantage of using Variable Projection to segregate the linear parameters from the nonlinear ones lies in simultaneous fitting without the need for a fixed threshold to split b-values for perfusion and diffusion components.

While fitting the volume fractions is a simple convex optimization problem, we still need a good estimate to fit the nonlinear exponent parameters. For this purpose, we make use of state-of-the art global optimization methods to get good initial estimates for the entangled exponential functions. We provide tools to efficiently switch between different types of global optimization: simplicial homology (SH) \citep{endres_sandrock_focke_2018} \citep{paulavicius_zilinskas_2014} and differential evolution (DE). Our results show that the SH and DE methods for global optimization give similar results, but SH reduces the number of function evaluations required and is therefore faster as compared to DE. The results obtained from either of the global optimizers and convex optimizer give a good initial estimate for the model parameters in their separated subspaces. We then make use of these initial estimates to fit the full IVIM model using the Trust Region Reflective optimizer via nonlinear least squares \citep{golub_pereyra_2003}.

\section{Methods}
\subsection{Overview of the fitting framework}
The proposed framework using variable projection \citep{golub_pereyra_2003, farooq_xu_nam_keefe_yacoub_georgiou_lenglet_2016} aims to disentangle the linear parameters (here, the associated perfusion and diffusion fractions) by projecting and solving the problem in a reduced subspace of the nonlinear parameters. This is done by taking an epigraphical projection of the measured signal $S$ onto the range of the $x_s$ which is assumed to have a closed form solution. The next step is to optimize for $x$ which is done via either the Simplicial Homology (SH) or Differential Evolution (DE) algorithm (explained in detail in Section \ref{sec:SH} and Section \ref{sec:DE} respectively). \\

Simplicial homology based topological optimization works by constructing homology groups of the hypersurface of the objective function. The SH optimizer is particularly useful as it is non-stochastic in nature in comparison with evolutional algorithms. The calculated homology groups perform optimization in a derivative-free setting \citep{rios_sahinidis_2012} so that heuristics to switch between the local and global search space are not needed. SH has not only shown to have comparative comparisons with DE and Basinhopping \citep{endres_sandrock_focke_2018}, but also gives the user a unique capability to track the optimization process via the constructed homology groups. In our hands, SH gives the same results as the DE optimizer with an improved speed of searching the hypersurface of the objective function.\\

Evolutionary computation is known to be particularly useful for fitting exponential functions. In the proposed framework, DE was chosen as the stochastic optimizer as it is known to outperform other metaheuristic optimization methods, such as Particle Swarm Optimization, Genetic Algorithms (GA), Harmony Search (HS) and Seeker Optimization Algorithms (SOA) \citep{ismail_halim_2017}. The particular advantage of using DE as an optimizer is that it gives the same result even if the model parameters are varied which is useful to speed up the search process and to avoid parameter tuning under different settings. Due to low values in the standard deviation of the fit, the results obtained from this approach improve the predictability in fitting. However, our experiments depict that using SH is a better strategy than DE for this type of a problem. Comparisons of SH against DE has been done in greater detail in global optimization literature 
\citep{endres_sandrock_focke_2018}.\\

After the global optimization step, convex optimization is  used to get better initial estimates of the projected volume fractions. We know that the volume fractions are combined as a convex combination (i.e. $f_{\textrm{perfusion}} + f_{\textrm{diffusion}} = 1$). Thus a simple search for these linear parameters in the reduced subspace improves the quality of the fit and helps obtain refined estimates for the respective volume fractions for the last stage. The nonlinear least squares solver is used as the final step for finding the parameter estimates, but now with improved initial estimates of the space. We show that the parameters obtained from this estimation significantly improve the quality of fitting.

\begin{figure}
  \centering
  \includegraphics[width=160mm]{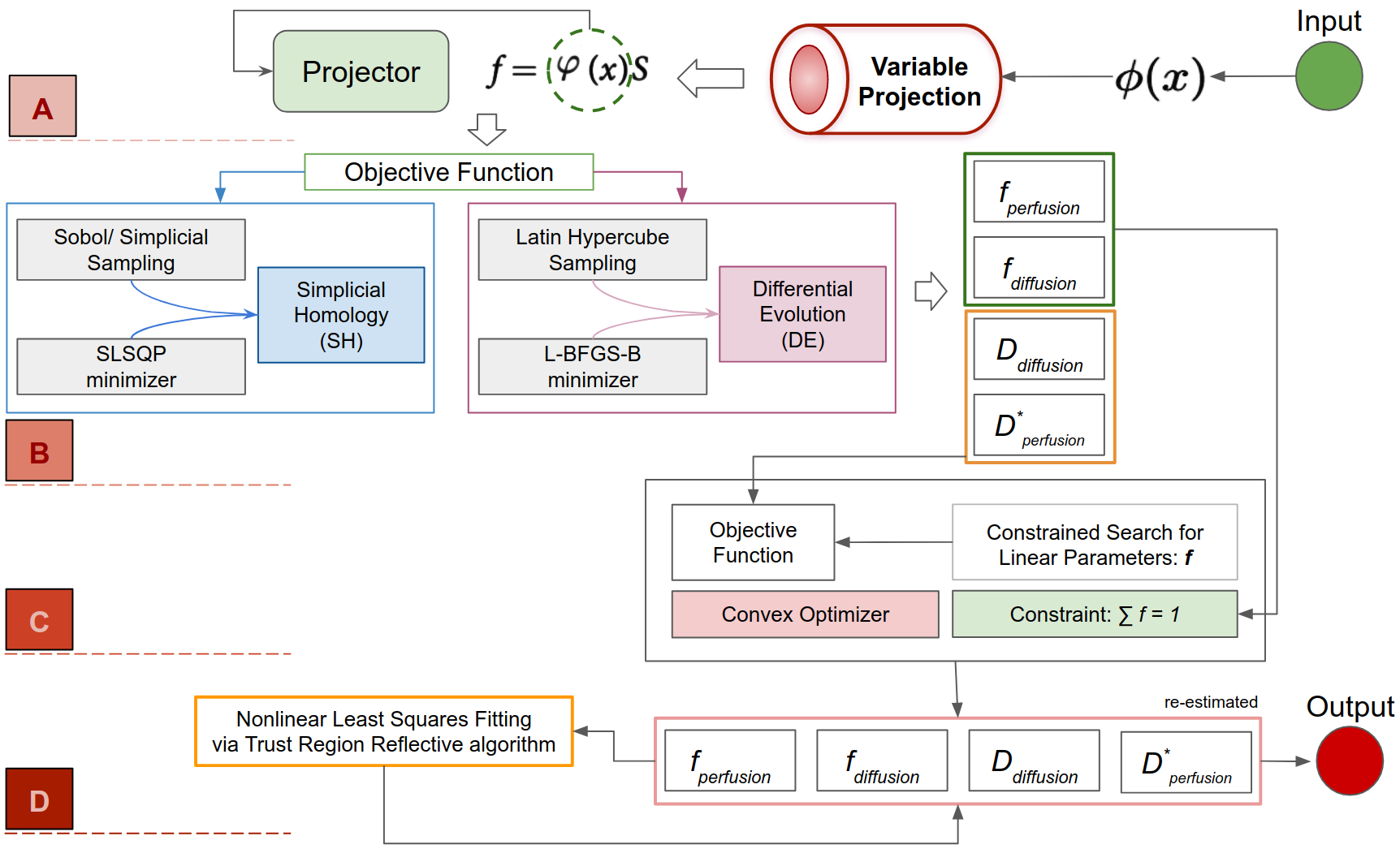}
    \caption{Schematic flow of the different stages of parameter estimation in the proposed framework: A. The Variable Projection step where the volume fractions are projected out and used in the consequent steps. B. The components of the simplicial homology (SH), differential evolution (DE) solver and the outputs it generates. C. The convex optimizer constrained by the volume fractions from the previous steps. D. Nonlinear least squares to re-estimate the final fit using the initial estimates obtained from the previous steps.}
\end{figure}

\subsection{Variable Projection (VarPro) for IVIM model}
The volume fraction parameters of IVIM model appear linearly, i.e. as linear combinations of variables that are nonlinearly parameterized (exponential components). Therefore, for every choice of the nonlinear parameters, we can solve for the linear coefficients $f$ as a convex optimization problem.

Problem formulation:
The IVIM model equation can be written as follows:
\begin{equation}
S = S_0\left(f_{ivim} e^{-b({D^* + D_{blood}})} + (1-f_{ivim})e^{-bD}\right)  
\label{eq:1}
\end{equation}

Here $f_\textrm{{ivim}}$ denotes the perfusion fraction and $1-f_\textrm{{ivim}}$ the fraction associated with diffusion component. From the model setup \citep{bihan_breton_lallemand_aubin_vignaud_laval-jeantet_1988}, we know that $\sum f_\textrm{{ivim}} = 1$. $D$ and $D^*$ are the nonlinear parameters that are entangled with the volume fractions. $D_\textrm{{blood}}$ denotes the diffusion coefficent of water in the blood. Note that the parameter $(D^* + D_\textrm{{blood}})$ together are just treated as $D^*$ during fitting as the fraction of $D_\textrm{{blood}}$ is much smaller as compared to the tissue water content.

To take the epigraphical projection of the measured signal, let us define the signal model as: $S{_1}(x) = e^{-bD^*}$ and $S{_2}(x) = e^{-bD}$ (isotropic models for the exponential decay of each of the individual components). Now we can define this set of nonlinear functions as our atom functions of a finite dictionary \citep{mitra_bhatia_2014} that we will combine as a linear sum of the volume fractions ($f$). We can set up our dictionary with the atom functions (here, signal models for perfusion and diffusion) as:
\begin{equation}
\phi(x) = \big[S{_1}(x), S{_2}(x)\big]
\label{eq:2}
\end{equation}
Thus,  from equations (\ref{eq:1}) and (\ref{eq:2}) we can write our objective function to fit the IVIM model as follows:
\begin{equation}
min_{x, f}\big\|S - \phi(x)f^{T}\big\|^2_2
\label{eq:3}
\end{equation}
Now, as mentioned \citep{golub_pereyra_2003}, we will project out the volume fractions with the help of a projector function. This is a Moore-Penrose inverse of the nonlinear components of our model defined in equation (2). We can write the projection function $\varphi$ as:
\begin{equation}
\varphi(x) = \left(\phi(x)^T\phi(x)\right)^{-1}\phi(x)^T    
\label{eq:4}
\end{equation}

We can make use of the measured signal and the projector function to now project out the linear components $f$ onto the range of $\phi(x)$ as:
$f = \varphi(x)S$. Using the above projected component for the volume fraction, we can reformulate the objective function of the IVIM model as in equation (\ref{eq:3}) where we only minimize over $x$ in a reduced subspace using the projector function described in equation (\ref{eq:4}). 
\begin{equation}
min_{x}\big\|S - \phi(x)\varphi(x)S\big\|^2_2
\end{equation}
This objective function is then used in the differential global optimization stage (via SH or DE) described in further detail in the sections (\ref{sec:SH} and \ref{sec:DE} respectively).

\subsection{Simplicial Homology Global Optimization}
The Simplicial Homology (SH) optimizer is a topographic optimizer that takes a derivative-free approach to optimization by constructing a simplicial homology over the surface of the objective function mentioned in equation (3). The SH attempts to find the quasi-equilibrium solutions to extract all the local minima from the sampled search space. While the algorithm guarantees convergence, it also provides the capability of tracking the progress of optimization and providing non-linear constraints to the objective function \citep{endres_sandrock_focke_2018}. \\

\begin{figure}
\centering
    \includegraphics[width=120mm]{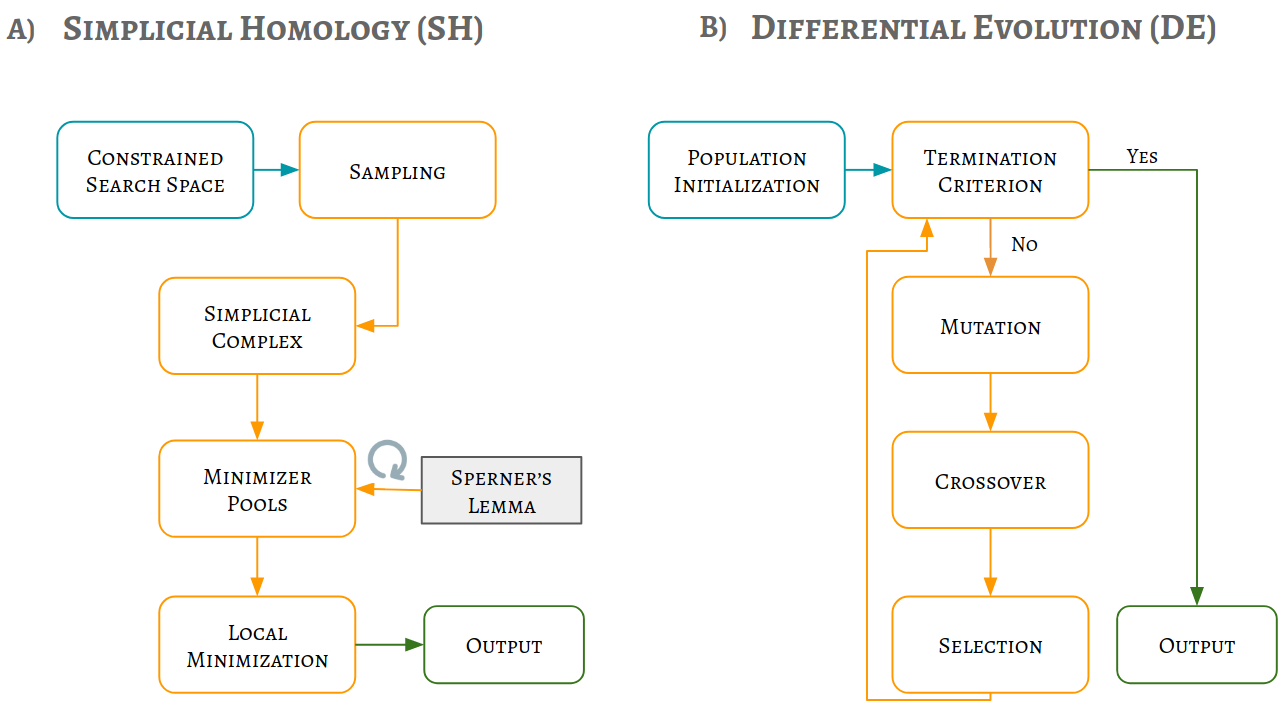}
    \caption{The basic steps involved in each of the algorithms used for Global Optimization: A) The basic steps of the workflow of the Simplicial Homology (SH) optimizer. B) The process of population generation and evolution in the Differential Evolution (DE) stochastic optimizer.}
    \label{fig:SHDE}
\end{figure}

The SH algorithm (depicted briefly in Fig. \ref{fig:SHDE}) can be divded into four basic phases: In the first phase, we generate the sample points in the constrained search space. These are the vertices of the simplicial complex to be constructed. In the second phase, the directed graph for the simplicial homology is constructed by triangulating these points. In the third phase, minimizer pools are generated from the mesh of simplicial complexes generated in the previous stage. This makes use of  Sperner's lemma \citep{bapat_1989} in a repeated manner to generate the pools for local minimization on top of the points sampled in the first phase from the objective function. In the last stage, local minimization is performed in each of the pools to thus reach a stationary global minimum. To perform the local minimization in each pool, we used the Sequential Least SQuares Programming (SLSQP) \citep{scipy_eric} \citep{golub_saunders_1969} optimization method.\\

One key aspect for setting up the SH optimizer is the sampling scheme to construct the homology groups from the objective function's hypersurface. For this purpose, we make use of the Sobol sequential sampling method \citep{sobol_1967} within the SH global optimizer. We notice that the SH optimizer also gives a significant speedup as compared to the DE optimizer. 
\label{sec:SH}\\

\subsection{Differential Evolution Optimization}
In the MIX approach to optimization, elitism based Genetic Algorithm (GA) was used as the optimizer for the nonlinear parameters. While GA are a very good approach to fit exponential functions, the stochastic nature of parameter tuning gives rise to a certain level of randomness in the fitting, which in turn affects reproducibility. We extend the previous work on MIX \citep{farooq_xu_nam_keefe_yacoub_georgiou_lenglet_2016} with an improvement in  fitting using simplicial homology (SH) and differential evolution (DE) algorithms \citep{price_storn_lampinen_2014}.\\

The IVIM model consists of exponential functions where parameters are hard to find \citep{istratov_vyvenko_1999}. To mitigate this difficulty, we make use of the latin hypercube initialization for the DE step. This improves sampling over the space, compared to random sampling. We make use of the 'best1bin' strategy within the DE optimizer, as defined in \citep{price_storn_lampinen_2014}. At each step of the DE, we make use of the Limited memory Broyden-Fletcher-–Goldfarb-–Shanno (L-BFGS-B) optimization \citep{byrd_peihuang_nocedal_1996} to improve the minimization process of the algorithm. The differential evolution this gives us an initial estimation of the model parameters: $f_\textrm{{ivim}}, D$ and $D^*$.
\label{sec:DE}

\subsection{Convex Optimization}
Based on these initial optimization steps, we have good initial estimates of the non-linear parameters $\phi(x)$. Using these initial estimates obtained from the DE, finding the $f$ is a convex problem and can be solved via a convex optimizer \citep{boyd_vandenberghe_2011} \citep{cvxpy} via constrained linear-least squares. The constraint imposed on the volume fractions is $\sum f_s =1$ and is used in the objective function of the convex optimizer as follows:
\begin{equation}
    min_f\big\|S - \phi(x)f^T\big\|_2^2 \ 
    subject\  to \  \sum f =1
\end{equation} 

\subsection{Nonlinear Least Squares Fitting}
As a final step, we make use of the initial estimates of the volume fractions and the nonlinear components as initialization for the full IVIM model. This ensures that the inter-dependency among the linear and nonlinear parameters is taken into account, with the final fit resulting in a stationary point. We make use of the Trust Region Reflective Algorithm \citep{branch_coleman_li_1999} to perform this type of optimization, due to its robustness and capability to handle sparsity in the data. It makes use of the reflective transformation and large scale approximation to find the optima of the search space \citep{scipy_eric}.

\begin{figure}
\centering
    \includegraphics[width=139mm]{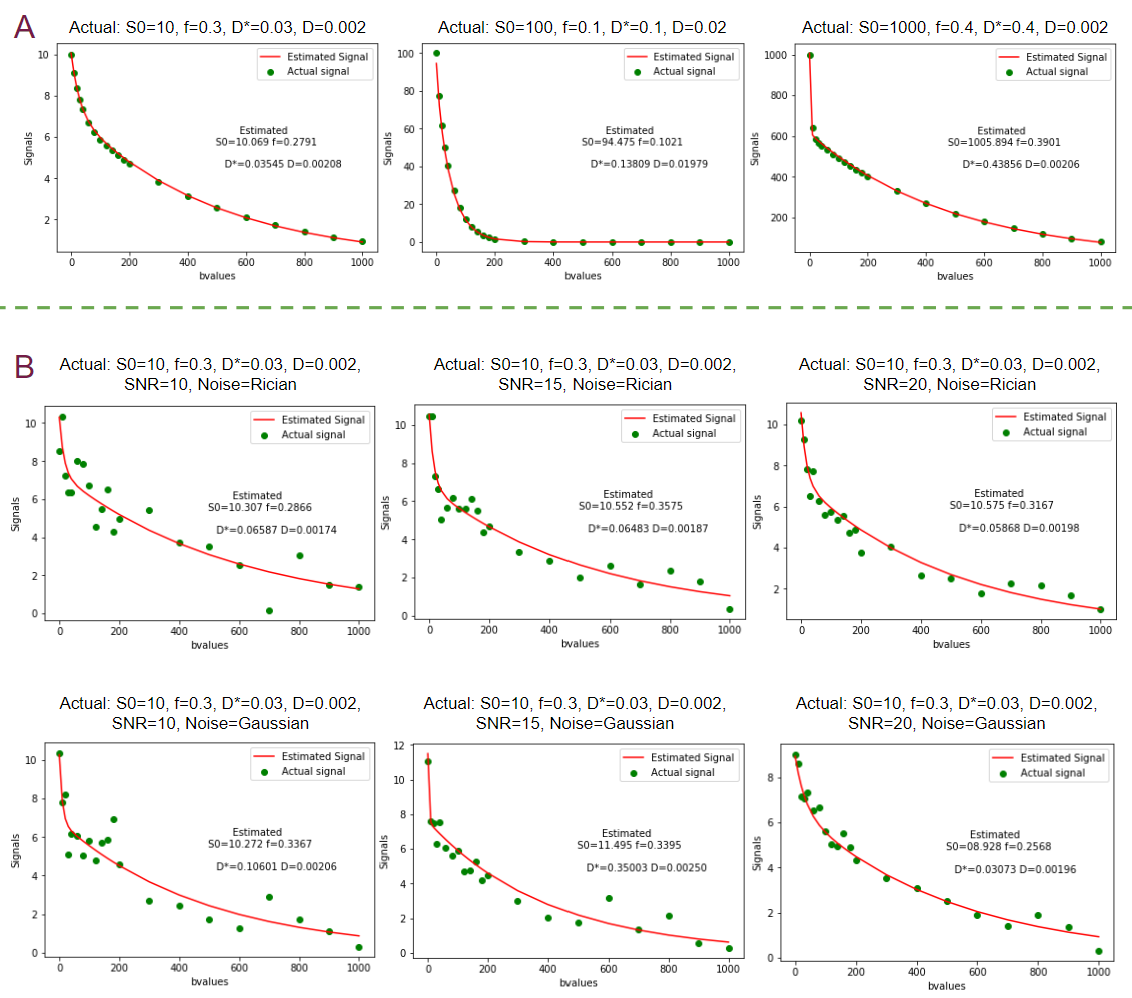}
    \caption{(A) Different parameter estimations based on simulations without noise. The parameters of the simulation are denoted at the top of each subplot. The parameter estimates are depicted beside the fitted curves. This figure shows that the model can fit various levels of data and a wide range of parameter values in an automatic and simultaneous estimation setting. (B) Different parameter estimations based on simulations with noise from the model. The input to the simulation is denoted at the top of each subplot and the estimations are depicted next to the fitted curves. The simulation parameters are kept the same, whereas the noise levels are varied across each row. The top row depicts fitting with Rician noise and the bottom row with Gaussian noise. Note that the varying levels of noise do not affect the fitting and give a good parameter estimation.}
    \label{fig:nonoise}
\end{figure}

\begin{figure}
    \centering
    \includegraphics[width=160mm]{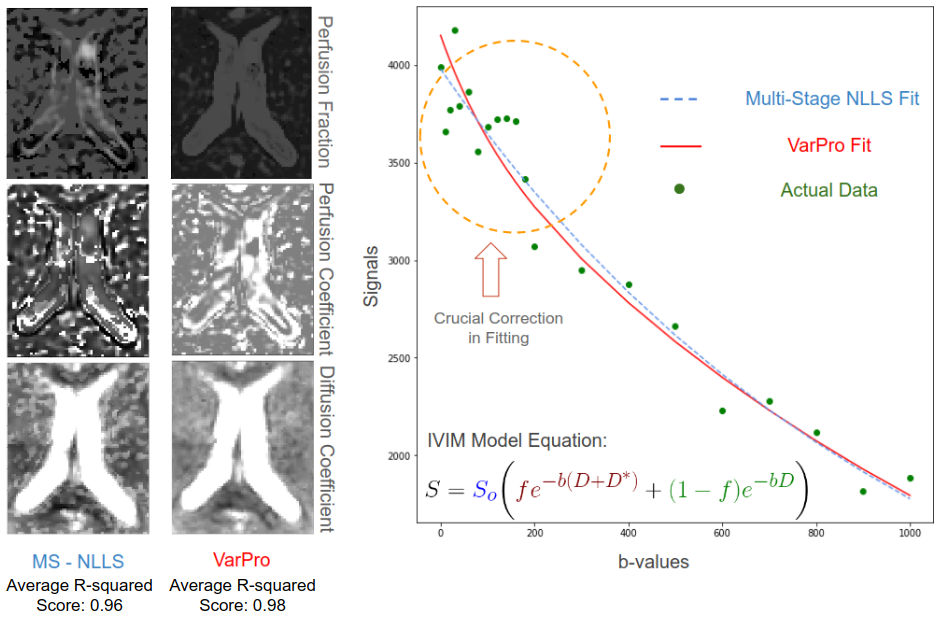}
    \caption{The difference between the generated maps by the MS-NLLS and the VarPro methods of fitting. Notice that the maps show clear segregation and a more interpretable contrast for different parameters. The right hand side of the image shows a better goodness-of-fit at lower b-values. We highlight the improvement in the generated contrast maps and the improved goodness-of-fit at lower b-values (IVIM effect typically occurs at b-values $<$ 400). The contrast maps provide more qualitative information via VarPro along with a better goodness of fit as compared with the Multi-Stage Nonlinear Least Squares Fit (MS-NLLS).}
    \label{fig:fit_gof}
\end{figure}

\begin{figure}
    \centering
    \includegraphics[width=160mm]{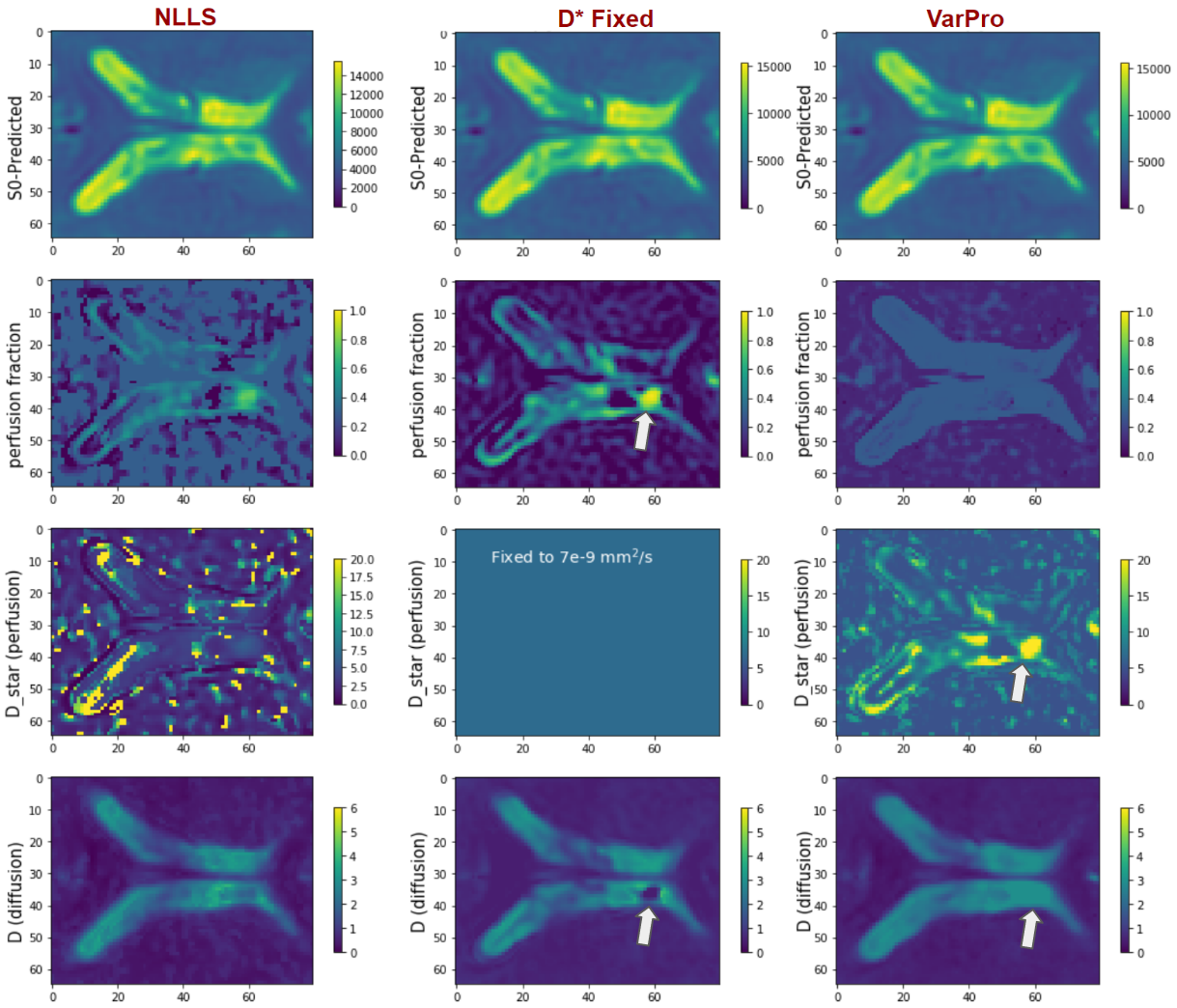}
    \caption{A comparison with Nonlinear Least Squares fitting, D* fixed Nonlinear Least Squares via \citep{fick_wassermann_deriche_2019} and Variable Projection frameworks. Row 1: the S0 signal prediction, Row 2: map for perfusion fraction prediction, Row 3: map for perfusion coefficient prediction and Row 4: map for diffusion coefficient prediction. We can clearly see that more qualitative information can be obtained from the VarPro fitting method (see section \ref{sec:results}). Notice that the perfusion and diffusion fractions have been clearly segregated without overfitting in the regions highlighted by the arrow.}
    \label{fig:maps}
\end{figure}
\label{sec:Methods}
\section{Experiments and Results}
\begin{figure}
    \centering
    \includegraphics[width=160mm]{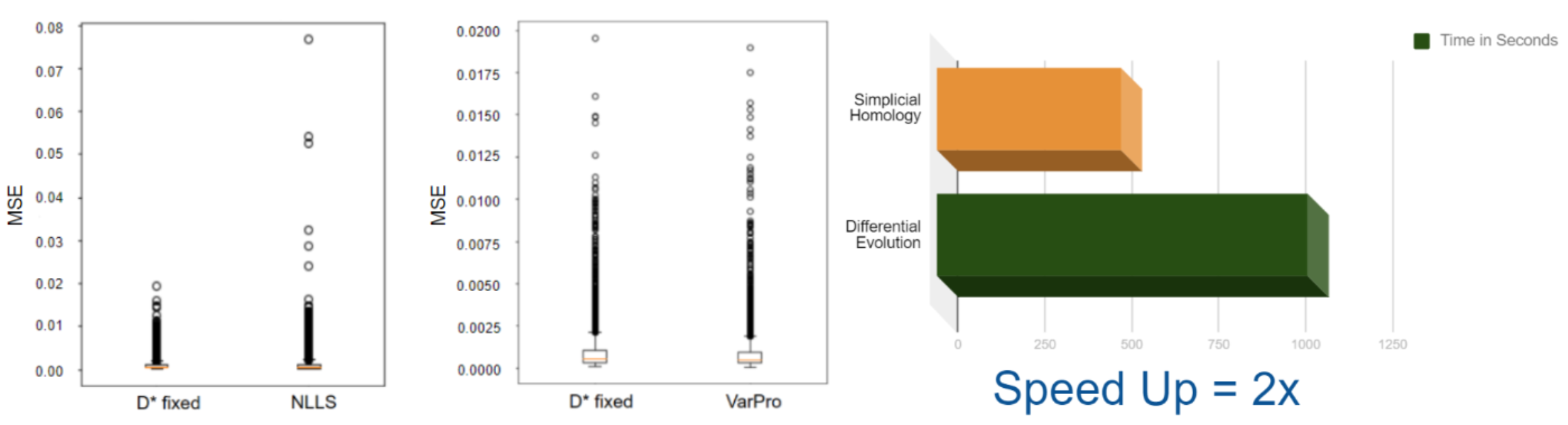}
    \caption{(i) Mean Squared Error (MSE) for the proposed method is lower. Box-plots of MSE in every voxel in the selected region of interest. Notice that the range of errors for the VarPro model is also much lower than the rest (NLLS and D* Fixed obtained from \citep{fick_wassermann_deriche_2019}). Note that the range of values was chosen differently for both the comparisons to show the variability in the $S0$ predictions for different settings. (ii) Shows the speed-up obtained from the Simplicial Homology (528sec) against Differential Evolution(1066sec) optimization (2x).}
    \label{fig:mse}
\end{figure}

The effort of this work has been to provide a good simultaneous fitting method via newer and more advanced optimization approaches. For the first time, we provide the Simplicial Homology (SH) optimization approach to improve the speed of global optimization process (2X speed-up shown in Fig. \ref{fig:mse}) initially proposed in MIX \citep{farooq_xu_nam_keefe_yacoub_georgiou_lenglet_2016}. To demonstrate that the proposed framework improves the quality of estimates amd to ensure the stability and robustness of fitting, we conducted experiments with both simulated (with and without noise) and real data \citep{peterson_2016} We simulated the signal from the IVIM model in DIPY \citep{garyfallidis_brett_amirbekian_rokem_walt_descoteaux_nimmo-smith__2014} and fitted it with the proposed framework. Fig. \ref{fig:nonoise} depicts different parameters used for simulating the data and their respective estimations. We can clearly see that the model gives almost exact estimates for the signal for each case. Additionally, signals have been simulated and fitted with both Rician and Gaussian noise at varying SNR level (Fig. \ref{fig:nonoise}), demonstrating the high quality of estimation and its robustness to noise. \\

Furthermore, we evaluate the goodness-of-fit per voxel using a 2-fold cross validation procedure \citep{Rokem_2015} and quantified via a $R^2$ (coefficient of determination) statistic. To do so, we fit the IVIM model to a part of the data (learning set) and then use the model to predict a held-out set (test set). Prediction of the held out data is done and recorded iteratively over the whole dataset. At the end of these iterations, a prediction of all of the data is compared directly to all of the data in that voxel. The $R^2$ score of the proposed method is always higher compared against other fitting methods. \\

Additionally, in Fig. \ref{fig:mse} we show that the results obtained from the VarPro estimation have a lower MSE in predicting the $S0$ as compared to the pervasive multi-stage NLLS method with and without fixing the $D^*$ parameter. It has also been shown in the literature that for the liver and pancreas \citep{gurney-champion_klaassen_2018}, setting the perfusion to $7*10^{-9}mm^2/s$ improves the quality of fitting. We compare our results against the same and show that MSE obtained from the model fitting for the predicted S0 is lower, with the advantage of not having to set any empirical constraints.

As demonstrated in Fig.~\ref{fig:maps}, when we fix the pseudo-diffusion/ perfusion parameter, the maps may end up with an artifact, as the estimation cannot find values for regions that go out of bounds (see comparison in bottom row). Furthermore, the figure also demonstrates that our method improves the estimation of the perfusion and diffusion fractions with more informative contrast maps. This is primarily due to the correct estimation of the associated fractions of perfusion and diffusion. VarPro fitting finds a good segregation between the parameter subspaces. In Fig.~\ref{fig:fit_gof}, we also show the goodness-of-fit at lower b-values, where the IVIM perfusion effect is more apparent.  
\label{sec:results}

In conclusion, we show that the proposed fitting method for parameter estimation is a promising method for IVIM. From Fig. \ref{fig:maps}, we see that the perfusion fraction is more interpretable, as compared to that obtained with other methods. We also see that the perfusion coefficient ($D^*$) has an agreement with the perfusion fraction obtained from fixing the $D^*$ to a certain value. While we can see that the MSE of the predicted $S0$ is lower than that of the other methods (see Fig. \ref{fig:mse}), the diffusion coefficient ($D$) is also properly estimated without artifacts from fixing other parameters. Therefore, we quantitatively and qualitatively show that the estimation is more precise than other methods.

\label{results}

\section{Discussion}
We provide new and improved strategies for estimating the IVIM parameters. For the first time, we make use of the Simplicial Homology Global Optimizer (SH) \citep{endres_sandrock_focke_2018} to speed up the fitting process. We provide means of model selection via k-fold cross validation and show that the results obtained from this fitting are better via the $R^2$ statistic. Our results show that the estimation maps obtained from the SH method of global optimization give the same results as DE and we therefore recommend using SH to gain speed performance. Our results show that the quality of estimation is better than other contemporary methods, and also overcomes the problem of bias that occurs in Bayesian \citep{while_2017} approaches through simultaneous fitting. While the proposed method has been applied and tested for IVIM, it can be easily extended to other bi-exponetial models in Diffusion MRI such as Free Water Diffusion Tensor Imaging (FW-DTI). Future work will extend this framework in DIPY to provide alternatives to fitting microstructural models via more advanced optimization and machine learning methods \citep{fadnavis_19}.

\section{Conclusion}
This paper introduces a novel method for IVIM model fitting via multistage optimization for automatic parameter estimation. The key features of this method include a combination of simplicial homology based topological global optimization and variable projection. Results show that the quality of estimation is improved with an increased accuracy in the fitting. 

\section{Acknowledgments}
NIH CRCNS grant R01EB027585 supported the work of Shreyas Fadnavis, Eleftherios Garyfallidis and Ariel Rokem. We would also like to thank the DIPY community (https://dipy.org) for distributing the method and providing meaningful feedback.

\bibliographystyle{apalike}
\small{{\center{\bibliography{ivim_varpro}}}}

\end{document}